\documentclass[manuscript]{aastex}
\usepackage{color}

\slugcomment{Not to appear in Nonlearned J., 45.}

\shorttitle{A detailed mass distribution of a high-density core}
\shortauthors{Tokuda et al.}

\begin{document}


\title{Revealing a detailed mass distribution of a high-density core MC27/L1521F in Taurus with ALMA}


\author{Kazuki Tokuda\altaffilmark{1}, Toshikazu Onishi\altaffilmark{1}, Tomoaki Matsumoto\altaffilmark{2}, Kazuya Saigo\altaffilmark{1}, Akiko Kawamura\altaffilmark{3}, Yasuo Fukui\altaffilmark{4}, Shu-ichiro Inutsuka\altaffilmark{4}, Masahiro N. Machida\altaffilmark{5}, Kengo Tomida\altaffilmark{6}, Kengo Tachihara\altaffilmark{4}, and Philippe Andr\'e\altaffilmark{7}}



\email{}

\altaffiltext{1}{Department of Physical Science, Graduate School of Science, Osaka Prefecture University, 1-1 Gakuen-cho, Naka-ku, Sakai, Osaka 599-8531, Japan}
\altaffiltext{2}{Faculty of Humanity and Environment, Hosei University, Fujimi, Chiyoda-ku, Tokyo 102-8160, Japan}
\altaffiltext{3}{National Astronomical Observatory of Japan, Mitaka, Tokyo 181-8588, Japan}
\altaffiltext{4}{Department of Astrophysics, Nagoya University, Chikusa-ku, Nagoya 464-8602, Japan}
\altaffiltext{5}{Department of Earth and Planetary Sciences, Kyushu University, Fukuoka 819-0395, Japan}
\altaffiltext{6}{Department of Earth and Space Science, Graduate School of Science, Osaka University, 1-1 Machikaneyama, Toyonaka, Osaka 560-0043, Japan}
\altaffiltext{7}{Laboratoire AIM, CEA/DSM-CNRS-Universit\'e Paris Diderot, IRFU/Service d'Astrophysique, C.E. Saclay, Orme des Merisiers, F-91191, Gif-sur-Yvette, France}


\begin{abstract}
We present the results of ALMA observations of dust continuum emission and molecular rotational lines toward a dense core MC27 (aka L1521F) in Taurus, which is considered to be at a very early stage of star formation.
The detailed column density distribution on size scales from a few tens AU to $\sim$10,000 AU scale are revealed by combining the ALMA (12 m array + 7 m array) data with the published/unpublished single-dish data.
The high angular resolution observations at 0.87 mm with a synthesized beam size of $\sim$0\farcs74 $\times$ 0\farcs32 reveal that a protostellar source, MMS-1, is not spatially resolved and lacks associated gas emission, while a starless high-density core, MMS-2, has substructures both in dust and molecular emission.
The averaged radial column density distribution of the inner part of MC27/L1521F ($r$ $\lesssim$ 3000 AU) is $N_{\mathrm{H_2}}$ $\sim$$r^{-0.4}$, clearly flatter than that of the outer part, $\sim$$r^{-1.0}$. The complex velocity/spatial structure obtained with previous ALMA observations is located inside the inner flatter region, which may reflect the dynamical status of the dense core. 

\end{abstract}


\section{Introduction}

\ Observational studies of the earliest stage of star formation are important for understanding their initial conditions, e.g., how a dense core fragments and collapses.
Recent observations suggest that individual protostellar collapse often leads to multiple protostellar systems \citep[e.g.,][]{Chen13}, although not as often as in purely hydrodynamic simulations, and confusion by small-scale structure in the associated protostellar outflow(s) is sometimes an issue \citep{Maury10}.
Complex gas structures have also been observed in the envelopes on spatial scales from 0.1 pc to $\sim$1000 AU \citep{Tobin10b,Tobin11}.  It is thus of vital importance to reveal the fragmentation process of a dense core to form multiple stars therein.  In spite of a number of observational efforts made thus far, some of the most fundamental issues are still left without good answers. Especially, an observational missing link between the complex envelope and formed protostars has still remained.  The observations revealing how the surrounding gas interacts to form stars at the center of a dense core are a key understanding the origin of binary or multiple systems.  Previous interferometric observations with high angular resolutions failed to detect extended high-density gas around protostars due to the lack of the sensitivity and the narrow spatial frequency coverage \citep[e.g.,][]{Maury10,Chen13,Takahashi13}.  The recent ALMA observations have been revealing the detailed structure of the innermost region of dense cores where multiple protostars are currently forming.\\
\ MC27 \citep{Mizuno94,Onishi96,Onishi98,Onishi99,Onishi02} or L1521F \citep{Codella97} is one of the densest cores in Taurus. It was considered as a prestellar core, until $Spitzer$ observations \citep{Bourke06} have detected a very low luminosity ($<$0.07 $L_{\odot}$) protostar (hereafter the $Spitzer$ source). It is of very high density and contains the $Spitzer$ source, which indicates it is among the youngest known Class 0 protostars \citep[cf.][]{Andre00} and still may preserve the initial conditions of star formation \citep{Bourke06,Terebey09}. \citet{Tokuda14} recently carried out ALMA Cycle 0 observations toward the object at an angular resolution of $\sim$1\arcsec\ by using the 12 m array, and revealing that the spatial and velocity distributions are very complex.  We detected (1) a few starless high-density condensations ($\sim$10$^{7}$ cm$^{-3}$), within a region of a several hundred AU around the $Spitzer$ source; (2) a very compact bipolar outflow centered at the $Spitzer$ source with a dynamical time of a few hundred years with an indication of interaction of surrounding gas; and (3) a well-defined long arc-like structure whose size is $\sim$2000 AU.  Subsequent numerical simulations demonstrate that a dense core undergoes gravitational collapse to form multiple protostars, and gravitational torque from orbiting protostars creates arc structures extending up to a thousand AU scale \citep{Matsumoto15}. These studies have been suggested that dynamical gas interactions are a key to forming the multiple (or binary) protostars.\\
 \ In spite of these new findings with ALMA, there is a gap in the coverage of the spatial frequencies between single dish observations with large aperture telescopes and the 12 m array of ALMA.  Therefore, additional investigations were really needed to understand the mechanism creating the above-mentioned complex structures and how protostars form in such an environment. \\
 \ One of the best methods to diagnose the evolutionary status of dense cores is to investigate the mass distribution because it should regulate the dynamics of cores. In Taurus, \citet{Onishi02} suggested that the peak density of 10$^{6}$ cm$^{-3}$, by measuring with a beam size of $\sim$20\arcsec, would be a threshold for the dynamical collapse. When the first core is formed, the density should continuously increase to the center, probably in the form of $n_\mathrm{H_2}$ $\sim$$r^{-2}$ \citep[e.g.,][]{Larson69,Masunaga98}. The finest beam size that can be achieved by single-dish telescopes is currently $\sim$11\arcsec\ by MAMBO-2 \citep{Kreysa99} or a new-generation continuum camera ``NIKA2''\citep{Calvo16} observations with the IRAM 30-m telescope, and the ALMA 12 m array cannot cover well the angular extent above several arcseconds.
 The mass distribution inside the beam size of $\sim$10\arcsec, corresponding to a spatial scale of 1400 AU at a distance of Taurus \citep[140 pc; e.g.,][]{Elias78}, is very important for determining the evolution because most of the stellar mass is included in the radius. Recently, an approach that combines data obtained with single-dish telescopes and interferometers has been widely used to investigate the physical properties in protostellar systems \citep[e.g., B335;][]{Yen11,Kurono13}.  ALMA has the capability to obtain the shorter spatial frequencies filling the gap between the single-dish and the 12 m array by using the ACA (Atacama Compact Array; Morita array). \\
\ In this paper, we present the results of higher-resolution observations in frequency Band 7 with ALMA to reveal the detailed structure of the central part of the core.  We then show the dust mass distribution, derived from 1.2 mm continuum observations, continuously from $\sim$100 AU to $\sim$10,000 AU by combining the data taken with the ALMA 12 m array, ALMA ACA 7 m array, and the IRAM 30 m/MAMBO-2.

\section{Observations}
\ We carried out ALMA Cycle 1 Band 6 (1.2 mm) and Band 7 (0.87 mm) continuum and molecular line observations toward MC27/L1521F at the center position of ($\alpha_{J2000.0}$, $\delta_{J2000.0}$) = (4$^{\rm h}$28$^{\rm m}$39\fs00, +26\arcdeg51\arcmin35\farcs0) with the 12 m array and the 7 m array (ACA).  
The observations have been carried out between 2013 October and 2015 June. 
There are four spectral windows for each Band, and the correlator for each spectral window was set to have a bandwidth of 1875 MHz with 3840 channels.  The total bandwidth is thus 7.5 GHz in each Band.  The Band 6 observations include, e.g., CS ($J$ = 5--4) and H$^{13}$CO$^{+}$ ($J$ = 3--2), and those of Band 7 include, e.g., C$^{17}$O ($J$ = 3--2), $^{12}$CO ($J$ = 3--2) and H$^{13}$CO$^{+}$ ($J$ = 4--3).
The $uv$ ranges of the Band 6 and the band 7 with the 12 m array data are 10.9--239 k$\lambda$ and 16.2--699 k$\lambda$, respectively. The 7 m array data of the Band 6 and the band 7 cover the baseline ranges of 8.1--27.7 k$\lambda$ and 8.9--43.9 k$\lambda$, respectively. The calibration of the complex gains was carried out through observations of four quasars (J0423-0120, J0510-0159, J0532+0732, and J0238+1636), phase calibration of two quasars (J0510+1800, and J0438+3004), and flux calibration of two solar system objects (Ganymede and Uranus) and a quasar (J0510+1800). For the flux calibration of the solar system objects, we used the Butler-JPL-Horizons 2012 model\footnote[1]{https:$//$science.nrao.edu$/$facilities$/$alma$/$aboutALMA$/$Technology$/$ALMA$\_$Memo$\_$Series\\$/$alma594$/$abs594}. 
The data were reduced using the Common Astronomy Software Application package\footnote[2]{http:$//$casa.nrao.edu} and the visibility imaged. We used the natural weighting for both the Band 6 and the Band 7 data. The synthesized beams and the sensitivities of the dust continuum and the molecular lines are listed in Table \ref{tablePram}. \\
\ The single-dish continuum observations of the 1.2 mm thermal dust emission were done between 2002 December and 2003 February with the IRAM 30 m telescope on Pico Veleta (Spain) as part of project 109--02 (P.I.: Ph. Andr\'e), using the 117-channel MAMBO-2 bolometer camera (240$\farcs$ diameter) of the Max-Planck-Institute f\"ur Radioastronomie \citep{Kreysa99}. Three on-the-fly maps of size $\sim6\arcmin \times 5\arcmin$ were obtained with this camera using a dual-beam raster mode and a scanning speed of $5\arcsec $ sec$^{-1}$ to $8\arcsec $ sec$^{-1}$ in azimuth. The wobbler frequency was 2 Hz and the wobbler throw (in azimuth) ranged from $45\arcsec$ to $60\arcsec$. The atmospheric optical depth at zenith, monitored by skydips, was between $\sim0.15$ and $\sim0.3$. The total integration time was about 2~hr, excluding calibration overheads, and the resulting rms noise level was about 3.7 mJy beam$^{-1}$ at the center of the final combined map of MC27/L1521F (cf. Figure \ref{Band6_Combine} (a)). The FWHM size of the main beam on the sky was measured to be $11\arcsec $ based on beam maps of Uranus. The central effective frequency was 250 GHz with half-sensitivity limits at 210 and 290 GHz. Pointing and focus position were usually checked before and after each map. We note that the similar observations of 1.2 mm continuum toward MC27/L1521F were carried out with the same instrument in previously published papers \citep{Crapsi04,Kauffmann08}, although our observation covers larger area and/or achieves higher sensitivity.

\section{Results}
\subsection{High-resolution observations of the central part of high-density dust cores}\label{High-res}
\ Figure \ref{MMS2} shows that the present 0.87 mm continuum observations detected three intensity peaks, and they correspond to MMS-1, MMS-2, and MMS-3 identified with the previous 1.2 mm continuum observations \citep{Tokuda14}. MMS-1, which coincides with the $Spitzer$ source, is not spatially resolved even with the current angular resolution of $\sim$0\farcs74 $\times$ 0\farcs34, which is consistent with the PdBI observations by \citet{Maury10}. 
The deconvolved size is (0\farcs28$\pm$0\farcs1) $\times$ (0\farcs13$\pm$0\farcs06), corresponding to (40$\pm$14 AU) $\times$ (18$\pm$8 AU). 
We have not detected significant molecular line emission enhancements toward MMS-1 in $^{12}$CO ($J$ = 3--2), H$^{13}$CO$^{+}$ ($J$ = 4--3), and C$^{17}$O ($J$ = 3--2), indicating that the $Spitzer$ source does not have extended gas envelopes. However, the dust emission and the compact outflow emitted from MMS-1 imply the existence of a disk-like structure around the central protostar, and thus further high angular resolution observations are definitely needed to investigate the nature and evolutionary stage of MMS-1.\\
\ The spatial distribution of the 0.87 mm continuum is similar to that of 1.2 mm, and the sizes of the whole extent of MMS-2 are almost the same, of $\sim$360 AU at 1.2 mm and 0.87 mm.  The 0.87 mm continuum observation revealed that MMS-2 could be further resolved into two peaks.  This fact is also supported by the spatial distribution of molecular line emissions of C$^{17}$O  ($J$ = 3--2).  
On the other hand, the spatial distribution of H$^{13}$CO$^{+}$  ($J$ = 4--3) emission shows the single peaked distribution.  This may be because the line is too optically thick to probe the central part in MMS-2; H$^{13}$CO$^{+}$  ($J$ = 3--2) spectra were also suggested to be optically thick \citep{Tokuda14}.  Actually, the integrated intensity ratio of H$^{13}$CO$^{+}$ ($J$ = 4--3)/H$^{13}$CO$^{+}$ ($J$ = 3--2) around the H$^{13}$CO$^{+}$ ($J$ = 4--3) peak in MMS-2 reached $\sim$0.8, indicating that the H$^{13}$CO$^{+}$ ($J$ = 4--3) is fully thermalized and that the region should be very high density of $>$10$^7$ cm$^{-3}$ with a calculation by a non-LTE analysis \citep[RADEX;][]{van07} under the assumption of the kinematic temperature of 10 K. \\
\ Another dust core, MMS-3, is located toward the northwest of the $Spitzer$ source with $\sim$3$\sigma$ detection. The C$^{17}$O ($J$ = 3--2) emission was not detected, but the H$^{13}$CO$^{+}$ ($J$ = 4--3) emission was detected toward this source.  The integrated intensity ratio of H$^{13}$CO$^{+}$ ($J$ = 4--3)/H$^{13}$CO$^{+}$ ($J$ =3--2) is $\sim$0.5, which corresponds to a density of $\sim$ 10$^6$ cm$^{-3}$ from the non-LTE calculation, and then the MMS-3 is less dense than MMS-2 by a factor of $\sim$10.  Nondetection of C$^{17}$O ($J$ = 3--2) is justified by the fact that the optically thin C$^{17}$O ($J$ = 3--2) traces the column density and that the column density derived from the continuum emission is less than half of the MMS-2 as shown below.\\
\ We derived the total mass and the column density from the 0.87 mm dust continuum data to investigate the physical properties of the condensations with the assumption of optically thin emission and uniform dust temperature. The column density of molecular hydrogen is estimated by using
\begin{equation}\label{eq_column}
N_\mathrm{H_2} = F^\mathrm{beam}_\nu/\Omega_\mathrm{A}\mu_\mathrm{H_2}m_\mathrm{H}\kappa_{\nu}B_{\nu}(T_\mathrm{d})
\end{equation}
and
\begin{equation}\label{eq_kappa}
\kappa_{\nu} =  \kappa_{\mathrm{231 GHz}}(\nu/\mathrm{231 GHz})^\beta
\end{equation}
where $F^\mathrm{beam}_\nu$ is the flux per beam at frequency  $\nu$, $\Omega_\mathrm{A}$ is the solid angle of the beam, $\mu_\mathrm{H_2}$ is the molecular weight per hydrogen, $m_\mathrm{H}$ is the H-atom mass, $\kappa_{\nu}$ is the mass absorption coefficient, $\kappa_{\mathrm{231GHz}}$ = 1.0 $\times$ 10$^{-2}$ cm$^2$\,g$^{-1}$ of interstellar matter is the emissivity of the dust continuum at 231 GHz \citep{Oss94,Kauffmann08}, $\beta$ is the dust emissivity index, $B_{\nu}$ is the Planck function, and $T_\mathrm{d}$ is the dust temperature. To estimate the total mass ($M_\mathrm{total}$), we use
\begin{equation}\label{eq_mass}
M_\mathrm{total} = F_{\nu}d^2/\kappa_{\nu}B_{\nu}(T_\mathrm{d})
\end{equation}
where $F_{\nu}$ is the integrated flux of the continuum emission at frequency $\nu$ and $d$ is the distance of the source (140 pc; Elias 1978).  We assumed $T_\mathrm{d}$ = 10 K, which is consistent with the spectral energy distribution of MC27/L1521F observed with $Herschel$ as part of the Gould Belt survey (\citealt{Andre10}; P. Palmeirim 2014 PhD thesis, private communication). $\beta$ was assumed to be 1.9, which is the average value derived from the ratios of 1.2 mm and 0.87 mm intensities ranging from 1.7 to 2.1 except toward MMS-1, and this value is consistent with those assumed for radiative transfer models of Class 0 envelopes \citep{Oss94} and other Class 0 observations (Shirley 2011).  The mass absorption coefficient at 0.87 mm is thus calculated to be $\kappa_{\mathrm{0.87 mm}}$=2.1 $\times$ 10$^{-2}$ cm$^2$\,g$^{-1}$. The derived parameters of the dust sources are shown in Table \ref{table1}. The parameters are consistent with the results derived from the 1.1 mm dust continuum data \citep{Tokuda14} within $\sim$10--20\% for MMS-2. 
A larger difference is seen toward MMS-3 by a factor of 2--3. This may be because the emission toward MMS-3 is almost a 3$\sigma$ detection in both frequencies, and thus the errors in the intensities are considered not to be small.\\
\ We note that recent $Herschel/SPIRE$ Fourier Transform Spectrometer (FTS) measurements toward the center ($\sim$30\arcsec) position of MC27/L1521F reported a dust temperature of 15.6$\pm$05 K and a $\beta$ of 0.8$\pm$0.1 \citep{Makiwa16}, which is different from what we assumed.  $\beta$ toward MMS-1 is calculated to be small ($\sim$0.4), with an angular extent of $\sim$1\farcs3 $\times$ 0\farcs8 in the present ALMA observations, and this fact may indicate that the inclusion of MMS-1 in a $Herschel$ beam may be a cause of the low beta and high temperature.  Therefore, we use $\beta$ = 1.9 and $T_{\rm d}$ = 10  K here when we derive the mass of the extended emission, and the mass is changed for different beta and $T_{\rm d}$ according to the equations (\ref{eq_column})--({\ref{eq_mass}).

\subsection{Detailed column density distribution toward MC27}\label{column_section}
Two figures in Figure \ref{Band6_7m12m} demonstrate the distributions of the 1.2 mm continuum observed by the 12 m array and 7 m array.  The 7 m array and IRAM 30-m data in Figure \ref{Band6_Combine} show the single peaked core centered around MMS-1 and MMS-2, and the 12 m array data resolved the complex structure at the innermost region.  The total flux obtained by the 12 m array observation in the observed area is $\sim$21 mJy, which is smaller than that by the 7 m array in the same area of $\sim$61 mJy, and even smaller than that of the IRAM 30 m data of $\sim$200 mJy.  This fact indicates that the 12 m array observations missed a significant amount of flux of the extended emission, and even the 7 m array observations have the same situation.  Therefore, we combined the three different datasets by using the feathering technique in an accordance with the CASA instruction obtained by ALMA; we cleaned the 7 m and 12 m data sets individually and combined the cleaned maps and the single-dish map in Fourier space.\\
\ Figure \ref{Band6_Combine}(a) shows the dust continuum image obtained by the IRAM 30 m/MAMBO-2. Figure \ref{Band6_Combine}(b) is the combined image of the three data set.  We obtained the column density with the equation (2) by assuming $T_d$= 10 K and $\kappa_{\nu}$ = 1.2 $\times$ 10$^{-2}$ cm$^{2}$\,g$^{-1}$ \citep{Oss94,Kauffmann08}.  The synthesized beam size is 1\farcs3 $\times$ 0\farcs8.  The complex structure is actually seen around the peak of the extended core emission.  The noise-like ragged structure seen in the envelope comes from the fact that the 7 m array and IRAM 30 m data set do not have sufficient sensitivities for the beam size.\\
\ Figure \ref{Band6_Band7_Combine} (a) and (b) show a comparison between the 1.2 mm and 0.87 mm dust continuum emissions. We used the JCMT/SCUBA data obtained by \citet{Shinnaga04} and \citet{Kirk05} as single-dish data. The combined image of the 0.87 mm were generated by the same technique as described above. Both Figure \ref{Band6_Band7_Combine} (a) and (b) roughly show the same spatial distributions, indicating that they are tracing the same gas/dust component. \\
\ Figure \ref{power_low} shows the mean radial profile of the H${_2}$ column densities centered at MMS-2 derived from the 1.2 mm observations. Even if we change the center to MMS-1, the change in the radial profile is very small and does not affect the discussion below.  
The radial profile derived from 0.87 mm continuum observations is quite similar to that from 1.2 mm observations. We here use only the 1.2 mm continuum data to investigate the mass distribution as the mapping size of the 1.2 mm continuum ALMA observations is larger.
The radial profile seems to be a combination of multiple slopes. The inner part has a slope of $\sim$$r^{-0.4}$, and the outer part has a steeper slope of $\sim$$r^{-1.3}$. We note here that the dual-beam on-the-fly mapping mode of the MAMBO-2, which was used for the present observations, suffers from a progressive loss of signal as angular radius increases \citep{Motte01}.  Therefore, the slope of the outer part is expected to be shallower than the observed radial profile.  A simulation, similar to that described in the Appendix of \citet{Motte01}, indicates that the observed radial profile is consistent with an intrinsic column density distribution $\sim$ $r^{-1.0\pm 0.2}$ beyond $\sim$3000 AU and up to $\sim$14,000 AU.
Therefore, if we assume a uniform dust temperature, spherically symmetric geometry, and optically thin dust emission, the steep slope of $N_\mathrm{H_2}$ $\sim$ $r^{-1.0}$ corresponds to a density distribution $n_\mathrm{H_2}$ $\sim$$r^{-2.0}$, which was observed in previous studies of MC27 \citep{Onishi99, Crapsi04, Tatematsu04} and actually ends below $\sim$3000 AU.\\
\ We note that we detected a 1.2 mm continuum peak toward  ($\alpha_{J2000.0}$, $\delta_{J2000.0}$) = (4$^{\rm h}$28$^{\rm m}$37\fs84, +26\arcdeg51\arcmin28\farcs4) (Figure \ref{Band6_7m12m}(b)).  
We detected the source (MMS-ex1 in Figure \ref{Band6_7m12m} (b)) both in Cycle 0 and Cycle 1 observations.
The total flux is measured to be $\sim$1 mJy 1\farcs3 $\times$ 0\farcs8 beam$^{-1}$, and there is no corresponding gas emission detected toward the peak.  One possibility is that the source is a young stellar object older than the protostar phase, although we cannot find corresponding stars in the $Spitzer$ and optical data.  Then, the most probable candidate is a distant external galaxy.  \\

\subsection{Distributions of surrounding gas traced by $^{12}$CO ($J$ = 3--2)}
\ The $^{12}$CO ($J$ = 3--2) line can be used to trace the lower-density gas compared with the other molecular lines in that we observed in Cycle 0 and Cycle 1.  Figure \ref{$^{12}$COchannel} shows the channel maps of the $^{12}$CO ($J$ = 3--2) observations with the 12 m array alone with contours of HCO$^{+}$ ($J$ = 3--2) of the Cycle 0 observations.  The $^{12}$CO ($J$ = 3--2) intensity map at 5 km s$^{-1}$ is tracing filamentary/core-like structures and is very similar to that in HCO$^{+}$ ($J$ = 3--2).  However, at 7 km s$^{-1}$, the $^{12}$CO ($J$ = 3--2) emission is not detected, although HCO$^{+}$ ($J$ = 3--2) shows the characteristic filamentary structure, which was shown in \citet{Tokuda14} as an indication of the dynamical interaction of dense gas.  This behavior can be explained by the extremely large optical depth of $^{12}$CO ($J$ = 3--2) compared with HCO$^{+}$ ($J$ = 3--2).  The spectral profile in black in Figure \ref{$^{12}$COoutflow}(b) is the averaged spectrum of the $^{12}$CO ($J$ = 3--2) over the region where the line is detected.  It is clear that the there is a clear dip at the velocities higher than $\sim$ 6 km s$^{-1}$, and this dip can be interpreted as the optical thickness of the line.  The intensity enhancement of HCO$^{+}$ ($J$ = 3--2) at 3 km s$^{-1}$ is a part of the outflow from the $Spitzer$ source \citep{Tokuda14}, and we obtained the $^{12}$CO ($J$ = 3--2) emission toward the region at the same velocity.  The profile in blue in Figure \ref{$^{12}$COoutflow}(b) is the averaged $^{12}$CO ($J$ = 3--2) spectrum toward the region, and there is a clear indication of the blueshifted component of the outflow even in $^{12}$CO ($J$ = 3--2).  Figure \ref{$^{12}$COoutflow}(a) shows that the distributions of the blue wing component are almost the same in both $^{12}$CO ($J$ = 3--2) and HCO$^{+}$ ($J$ = 3--2).  However, the $^{12}$CO ($J$ = 3--2) emission was not detected toward the region where we detected the red wing component of the bipolar outflow detected in HCO$^{+}$ ($J$ = 3--2) by \citet{Tokuda14}.  This may be due to the optical thickness of the $^{12}$CO ($J$ = 3--2) line at the velocity range.  Actually, the $^{12}$CO ($J$ = 3--2) intensity of 7--9 km s$^{-1}$ is very weak, as shown in Figure \ref{$^{12}$COoutflow}(b) although the single-dish observations revealed strong emission at the same velocity range \citep{Takahashi13}.\\

\section{Discussion}
\subsection{Density profile}
\ The continuum distributions and the detailed radial column density profiles were presented in Sec. \ref{column_section}. First, we will summarize the past studies for the radial (column) density profiles toward protostellar and/or prestellar cores based on the (sub)millimeter continuum observations and compare them with our present results toward MC27/L1521F. 
\ MC27/L1521F is regarded as a dense core at a protostellar stage.  However, the luminosity of the source is quite low \citep{Bourke06} and the envelope of the protostar has a large amount of gas, indicating that the system preserves the initial condition of the protostar formation. We here compare the density profiles of MC27/L1521F with the past observational studies of prestellar/protostellar cores to investigate the evolutionary status.\\
\ Previous (sub)millimeter continuum observations toward Class 0/I sources in Taurus \citep[e.g.,][]{Motte01,Chan00,Shirley00} have revealed that the density profiles of the protostellar cores show $\rho(r) \propto r^{-p}$ with $p$ $\sim$1.5--2.0 over $\sim$10,000--15,000 AU in radius. \citet{Ward07} mentioned that the protostellar envelopes are more centrally condensed than prestellar cores and do not show the inner flattering in their radial column density profiles. These results are roughly consistent with the predictions of the isolated star formation model \citep[e.g.,][]{Shu77}.  \citet{Kirk05} obtained similar results toward a number of ($\sim$30) prestellar cores and categorized them into two groups, `bright' cores and `intermediate' cores, based on their flux intensities of the submillimeter continuum emissions. They found that intermediate cores are in agreement with a Bonner--Ebert sphere \citep[see also][]{Alves01}. On the other hand, in bright cores, the critical Bonner--Ebert sphere is not consistent with the radial density profiles of the cores, suggesting that they are already collapsing, or there is some additional mechanism (e.g., magnetic field) to support the cores. MC27/L1521F was characterized as the bright core of their work. The flattering radius ($R_\mathrm{flat}$) measured with 850 $\mu$m data is 3400 AU.\\
\ Recent multiwavelength studies obtained by $Herschel$ space observatory \citep{Lau13} determined the radial profiles of the column density and the temperature toward both protostellar cores and prestelaller cores and imply that the column density profiles are flat toward the peak. \citet{Roy14} also discussed similar results toward two prestellar cores (B68 and L1689B).
High angular resolution observations with the interferometer combining the single-dish data have revealed that the density profiles of the Class 0 source follows multiple power laws. \citet{Harvey03a,Harvey03b} derived the density profile of B335 core, containing a confirmed Class 0 source with a broken power law with a shallow density profile inside $\sim$6500 AU, $n(r) \propto r^{-1.5}$ and $r^{-2.0}$ outside \citep[see also][]{Kurono13}.  \citet{Kurono13} mentioned that the profile follows the inside-out collapse model \citep[e.g.,][]{Shu77}.\\
\ The present ALMA data on MC27/L1521F show that the radial profile of the density seems to be a combination of at least two slopes: steeper outside and shallower inside as presented in section \ref{column_section}.  Inside 3000AU, the density slope is $n_\mathrm{{H_2}}$$\sim$$r^{-1.5}$, which mimics the slope that is expected in the region where the inside-out collapse is occurring, as is in the case of B335 above.  The inside-out collapse expands outward with a velocity of the sound speed of $\sim$0.2 km s$^{-1}$ at 10 K \citep{Shu77}, and thus it takes $\sim$7$\times$10$^4$ yr to reach 3000 AU.  
However, the extreme faintness of the infrared source implies that the infrared source is at the very early stage of protostar formation, much earlier than 7$\times$10$^4$ yr.  This fact is consistent with a quite short dynamical time of the outflow of a few hundreds years \citep{Tokuda14}.
Moreover, the observed gas velocity shown in Figure \ref{$^{12}$COchannel} indicate a complex structure rather than a coherent infall motion owing to the inside-out collapse.  Therefore, the shallow slope of the radial profile of the density is probably not attributed to the inside-out collapse in the case of MC27/L1521F.
\ A possible scenario is that the shallow slope of the radial density profile is a consequence of a dynamical interaction between the high-density condensations and the envelopes because the inner part having the shallow slope overlaps with the region with the complex structures, such as arc-like structures we observed. 
As shown in the channel maps of Figure \ref{$^{12}$COchannel}, the velocity of the complex structures ranges in $\sim \pm1\ \mathrm{km\,s}^{-1}$ with respective to the systemic velocity.  The time scale for disturbance to propagate to a 3000 AU scale is estimated to be $1.4 \times 10^4~\mathrm{yr} \,(=3000~\mathrm{AU}/1~\mathrm{km\,s}^{-1})$, which is consistent with the age of the MMS-1 protostar in the early phase.  \\
\ According to dust continuum emission obtained by combining the ALMA 12 m array and 7 m array with IRAM 30-m, the total mass of the gas within $r \lesssim 3000~\mathrm{AU}$ is estimated as $M_\mathrm{gas} = 0.49\,M_\odot$, yielding a typical dynamical velocity of $0.4\,\mathrm{km\,s}^{-1} \, (= [ G M_\mathrm{gas}  / 3000~\mathrm{AU} ]^{1/2})$.  
This velocity is significantly lower than the observed velocity of $1~\mathrm{km\,s}^{-1}$, and this implies that the gas flow on this scale is caused by a dynamical interaction on a smaller scale.  \citet{Matsumoto15} demonstrate that a dynamical interaction on a 100 AU scale disturbs the envelopes on a $\sim$1000 AU scale, reproducing the arc-like structure on that scale.\\
\ \cite{Tokuda14} suggested that the very high density core MMS-2 is one of the most evolved starless condensations, or may contain a very faint protostellar object, which is a strong candidate for the first protostellar core, the first quasi-hydrostatic object during the star formation process \citep[e.g.,][]{Larson69,Masunaga98,Tomida13}. The MMS-2 is further resolved into two peaks by the present high-angular resolution observations with 0.87 mm continuum, as demonstrated in Sec. \ref{High-res}. We may be witnessing a fragmentation of the high density core in an early phase of star formation. We need further high-angular resolution observations with the long-baseline ALMA configurations to reveal the actual evolutionary stage of the MMS-2.

\subsection{Envelope structure seen in $^{12}$CO ($J$ = 3--2)}
\ \citet{Takahashi13} claimed that they detected molecular outflows in the $^{12}$CO ($J$ = 2--1) line with the SMA; one is originated from the $Spitzer$ source, and another is from an unknown source.  They stated that the blueshifted CO emission is brighter on the eastern side, which is in the opposite sense to what might be expected from the reflection nebula image obtained by the $Spitzer$ observations \citep{Bourke06,Terebey09}. Their  speculation of the cause is a consequence of the non uniform distribution of the surrounding material.  On the other hand, \citet{Tokuda14} detected a compact outflow in HCO$^{+}$ ($J$ = 3--2) with ALMA, and the direction of the outflow is consistent with the $Spitzer$ reflection nebula.
In addition to this, we detected the blue wing component in the $^{12}$CO ($J$ = 3--2) in the present observations, which coincides with that in HCO$^{+}$ ($J$ = 3--2).  Here we check the SMA outflows with the present ALMA observations in $^{12}$CO ($J$ = 3--2) line.  It is clear that they have not detected the outflowing gas corresponding to the HCO$^{+}$ ($J$ = 3--2) outflow because the extent is quite small ($\sim$2\arcsec).  If we spatially smooth the angular resolution of the channel maps in Figure \ref{$^{12}$COchannel} to the SMA resolution of \citet{Takahashi13}, the $^{12}$CO ($J$ = 2--1) and $^{12}$CO ($J$ = 3--2) distributions are quite similar, indicating that both observations are tracing the same gas.  The blueshifted velocity component of Figure 2 of \citet{Takahashi13} is seen mainly at 5 km s$^{-1}$ in Figure \ref{$^{12}$COchannel}, and they are actually filaments/cores that are also seen in HCO$^{+}$ ($J$ = 3--2) with the velocity range from 4 to 7 km s$^{-1}$.  A similar structure is also seen in H$^{13}$CO$^{+}$ ($J$ = 3--2), as shown in Figure 1 of \citet{Tokuda14}.  Therefore, this blueshifted velocity component is not the high-velocity gas from the outflow, but the relatively high-density gas structure at the center of this object.  The redshifted velocity component seems to be a bit complicated.  The $^{12}$CO cloud at ($\alpha_{J2000.0}$, $\delta_{J2000.0}$) = (4$^{\rm h}$28$^{\rm m}$38\fs80, +26\arcdeg51\arcmin38\farcs3), indicated by a dashed orange rectangle in Figure \ref{$^{12}$COchannel} corresponds to the HCO$^{+}$ ($J$ = 3--2) cloud possibly interacting with the HCO$^{+}$ ($J$ = 3--2) outflow that is shown in red dashed contours in Figure 2 (b and c) of \citet{Tokuda14}. 
We thus conclude that the blueshifted and redshifted components in $^{12}$CO ($J$ = 2--1) in \citet{Takahashi13} are not representing the outflowing gas emitted from the protostars, but a part of the complex dense gas structure in the core.\\
\ We note that the scattered light seen in the $Spitzer$ is larger than the molecular outflow observed with ALMA  in HCO$^+$ ($J$ = 3--2) \citep{Tokuda14}. Our ALMA results show that the $^{12}$CO ($J$ = 3--2) observations also failed to trace the cavity structure.  Recently, some observational evidences toward other Class 0 objects have been reported that such conical-shaped scatted light cannot be explained only by outflow activities \citep[e.g.,][]{Tobin10a,Tobin11}. They suggest that the scattered light is most likely be a vertically extended disk-in-forming with material falling onto it, indicating that the pseudo-disk is formed around the protostar and the light from the protostar is escaping toward the direction perpendicular to the disk.  In this case, there is no need for the outflow to follow the cavity-like structure.  
In the Cycle 0 \citep{Tokuda14} and the present Cycle 1 observations, we cannot identify any molecular/dust structure corresponding to the cavity-like structure.  This may be because the molecular lines such as HCO$^{+}$ and H$^{13}$CO$^{+}$ ($J$ = 3--2) can only trace the higher density gas, the dust emission is contaminated with the dust distribution in the same line of sight, and the $^{12}$CO ($J$ = 3--2) suffers from the extremely high optical depth to trace the central region.
We need additional observations to trace the cavity-like structure in other lines that can trace lower density envelopes having relatively small critical density and small optical depth such as $^{13}$CO and C$^{18}$O lines to investigate the cavity structure of MC27/L1521F. 

\section{Summary}
\ We present new interferometric dust continuum and molecular line observations of the dense core MC27/L1521F, containing a very faint protostar made with the ALMA Cycle 1 at wavelengths of 1.1 and 0.87 mm, including the Atacama Compact Array (ACA, aka Morita Array). We combine these data with previous dust continuum observations obtained by single-dish telescopes. Our main results are summarized as follows.\\ \\
\ 1. High angular resolution ($\sim$0\farcs74 $\times$ 0\farcs32) observations with ALMA 0.87 mm dust continuum show that MMS-1, the faintest protostar (the $Spitzer$ source), is still not spatially resolved and MMS-2, a starless high-density ($>$10$^{7}$ cm$^{-3}$) core, is further resolved into two peaks. Further high angular resolution observations with the long-baseline ALMA configurations are needed to reveal the detailed evolutionary stages of these sources.\\
\ 2. Detailed column density distributions with the sizes from $\sim$100 to $\sim$10,000 AU scale are revealed by combining the 12 m array (Main array) data with the 7 m array (ACA) data, as well as with the single dish IRAM/MAMBO-2 or JCMT/SCUBA data. The spatial dynamic range is significantly improved by an order of magnitude compared to previous studies. Our analysis shows that the averaged radial column density distribution of the inner part ($r <$ 3000 AU) is $N_{\mathrm{H_2}}\sim r^{-0.4}$, clearly flatter than that of the outer part, $\sim$$r^{-1.0}$. We detected the complex structure presented by previous ALMA observations \citep{Tokuda14} inside the inner flatter region, which may reflect the dynamical status of the dense core.\\
\ 3. We compared our $^{12}$CO ($J$ = 3--2) data with previous $^{12}$CO ($J$ = 2--1) data taken by the SMA \citep{Takahashi13}. We concluded that the $^{12}$CO redshifted and blueshifted velocity components observed by \citet{Takahashi13} are not outflowing gases ejected from the protostellar sources but a part of the complex gas structure. We failed to trace the cavity-like structure seen in scattered light in both HCO$^+$ ($J$ = 3--2) and CO ($J$ = 3--2), which are tracers for outflow.  This implies that the structure was not created only by molecular outflow, and further observations tracing the envelope in optically thin lines would be needed to investigate the cavity-like structure seen in $Spitzer$.

\acknowledgments

\ This paper makes use of the following ALMA data: ADS/JAO.ALMA\#2011.0.00611.S and 2012.1.00239.S. ALMA is a partnership of the ESO, NSF, NINS, NRC, NSC, and ASIAA. The Joint ALMA Observatory is operated by the ESO, AUI$/$NRAO, and NAOJ. This work was financially supported by the Japan Society for the Promotion of Science (JSPS) KAKENHI grant nos. 16H02160, 23103005, 23244027, 22244014, 23403001, 23540270, 24244017, 26247026, 26287030, and 26400233.  K. Tokuda is supported by a JSPS Research Fellow. T. Onishi supported by the ALMA Japan Research Grant of NAOJ Chile Observatory, NAOJ-ALMA-0013 and NAOJ-ALMA-0032. Ph. Andr\'e is partly supported by the European Research Council under the European Union'{}s Seventh Framework Programme (ERC Grant Agreement no. 291294). This work is partly based on observations carried out with the IRAM 30 m Telescope. IRAM is supported by INSU/CNRS (France), MPG (Germany), and IGN (Spain).

\clearpage

\begin{figure*}
 \begin{center}
  \includegraphics[width=160mm]{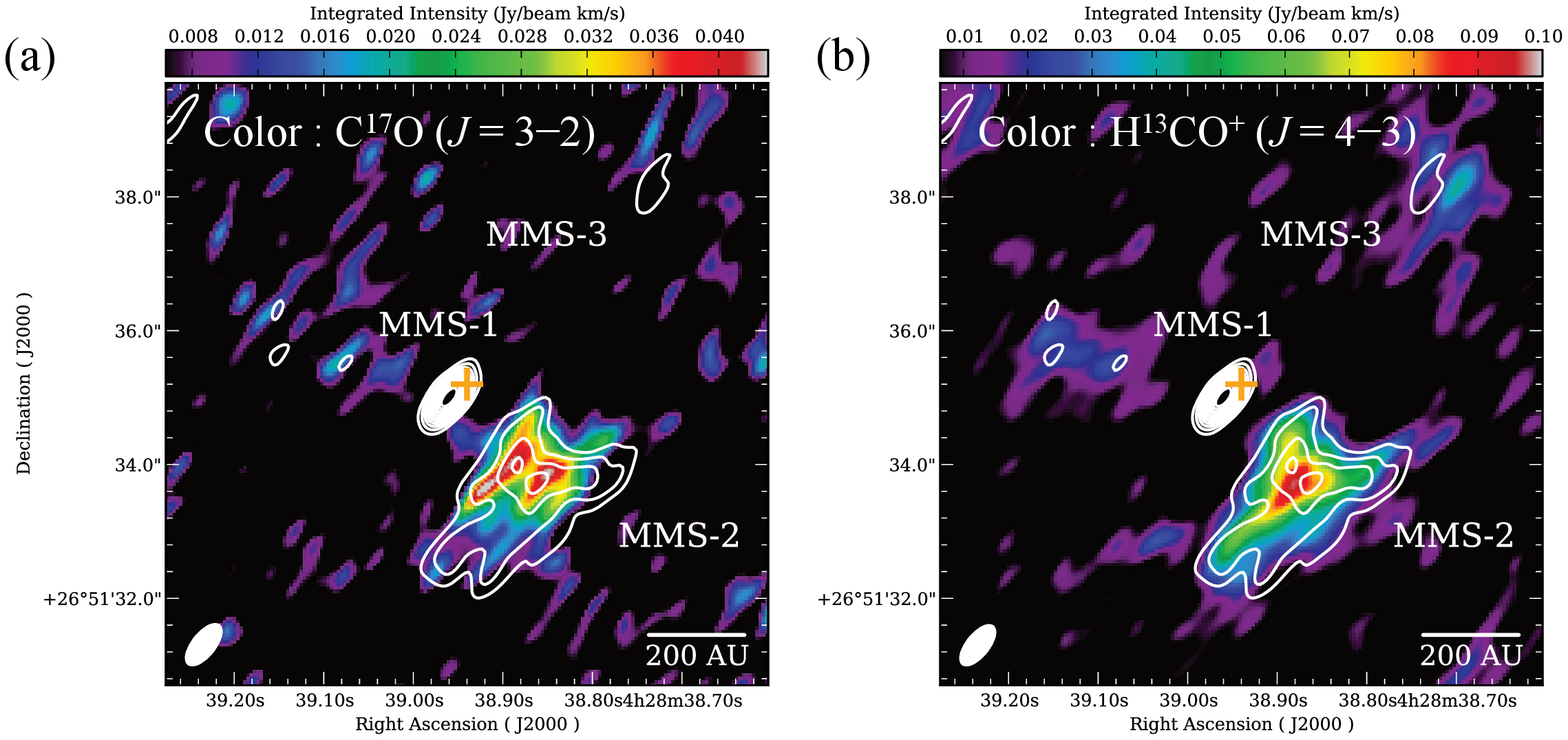}
 \end{center}
 \caption{Gas/dust distributions of the high-density condensations of MC27/L1521F in Band 7 obtained by the 12 m array alone.  (a) Total velocity-integrated intensities of C$^{17}$O  ($J$ = 3--2) are shown in color scale. (b) Total velocity-integrated intensities of H$^{13}$CO$^{+}$  ($J$ = 4--3) are shown in color scale. White contours represent the image of the 0.87 mm dust continuum emission in both (a) and (b). 
 The lowest contour and subsequent contour step are 3$\sigma$ and 2$\sigma$; 1$\sigma$ $\sim$0.1 mJy beam$^{-1}$.
 The angular resolutions are given by the ellipses in the lower left corners in each panel, 0\farcs73 $\times$ 0\farcs34. Orange crosses in each panel represent the position of the $Spitzer$ source.}
 \label{MMS2}
\end{figure*}

\begin{figure*}
 \begin{center}
  \includegraphics[width=160mm]{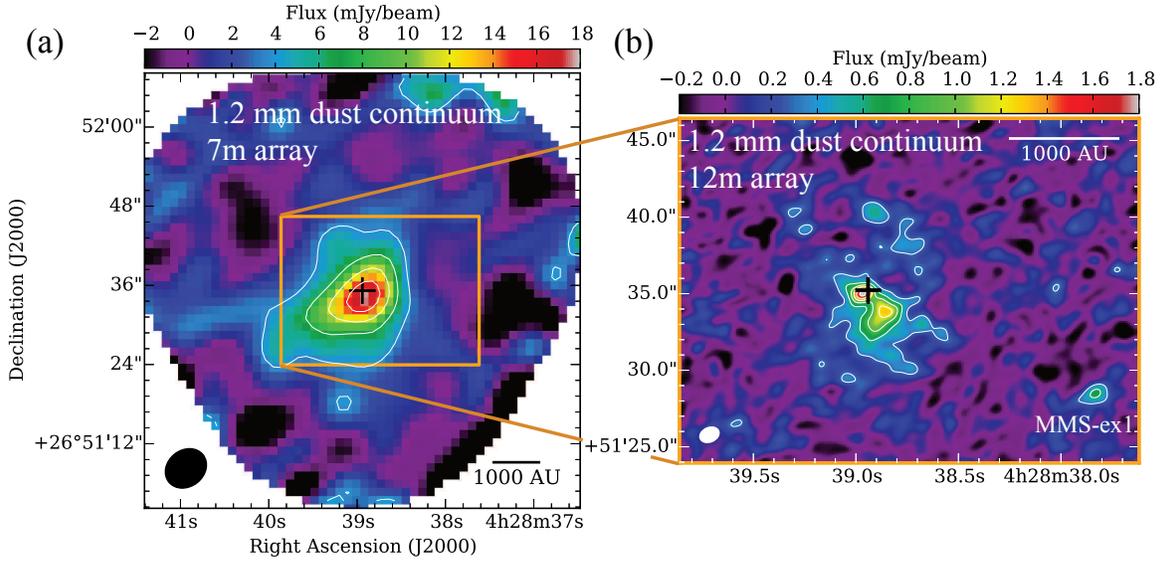}
  \end{center}
 \caption{Distribution of  1.2 mm dust continuum emission toward MC27/L1521F. (a) Both color scale and contours show 1.2 mm dust continuum images obtained by the 7 m array. The lowest contour and subsequent contour step are 3.6 mJy beam$^{-1}$. (b) Both color scale and contours show 1.2 mm dust continuum images obtained by the 12 m array The lowest contour and subsequent contour step are 0.3 mJy beam$^{-1}$. The angular resolutions are given by ellipses in the lower left corner of each panel. Black crosses in each panel represent the position of the $Spitzer$ source.}
 \label{Band6_7m12m}
\end{figure*}

\begin{figure*}
 \begin{center}
  \includegraphics[width=160mm]{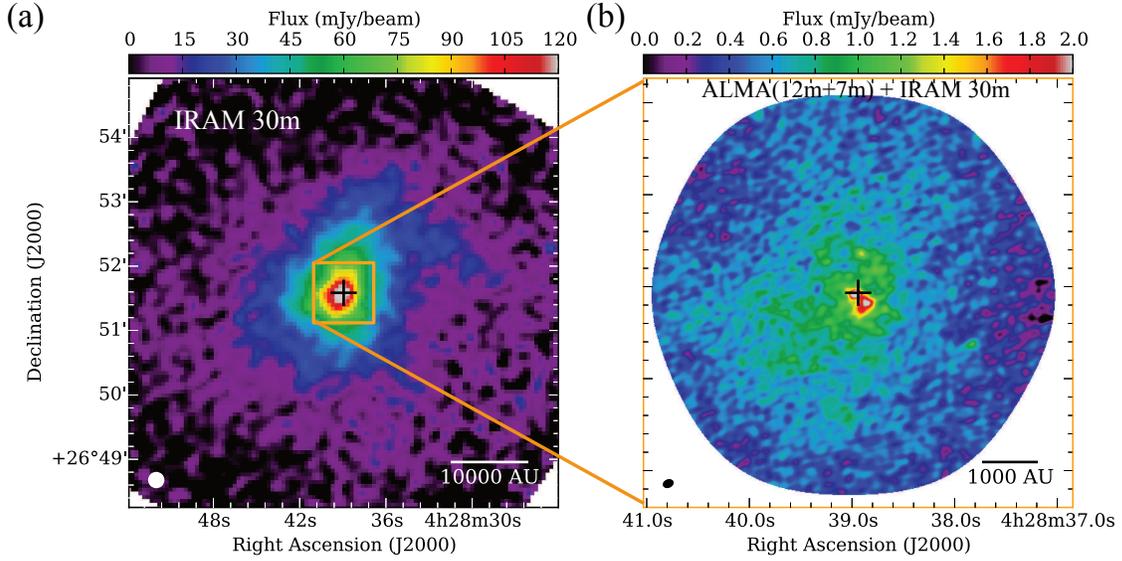}
 \end{center}
 \caption{Distributions of 1.2 mm dust continuum emission toward MC27/L1521F. (a) Color scale shows 1.2 mm dust continuum image with the IRAM 30 m telescope using the MAMBO-2 bolometer at angular resolution of 14\arcsec. The angular resolution is given by the circle in the lower left corner. (b) Color scale image shows the 1.2 mm dust continuum image combining the ALMA data (the 12 m array + the 7 m array) with IRAM 30-m data. The angular resolution of the combined image is given by black ellipse in the lower left corner, 1\farcs3 $\times$ 0\farcs8. Black crosses in each panel represent the position of the $Spitzer$ source.}
 \label{Band6_Combine}
\end{figure*}

\begin{figure*}
 \begin{center}
  \includegraphics[width=160mm]{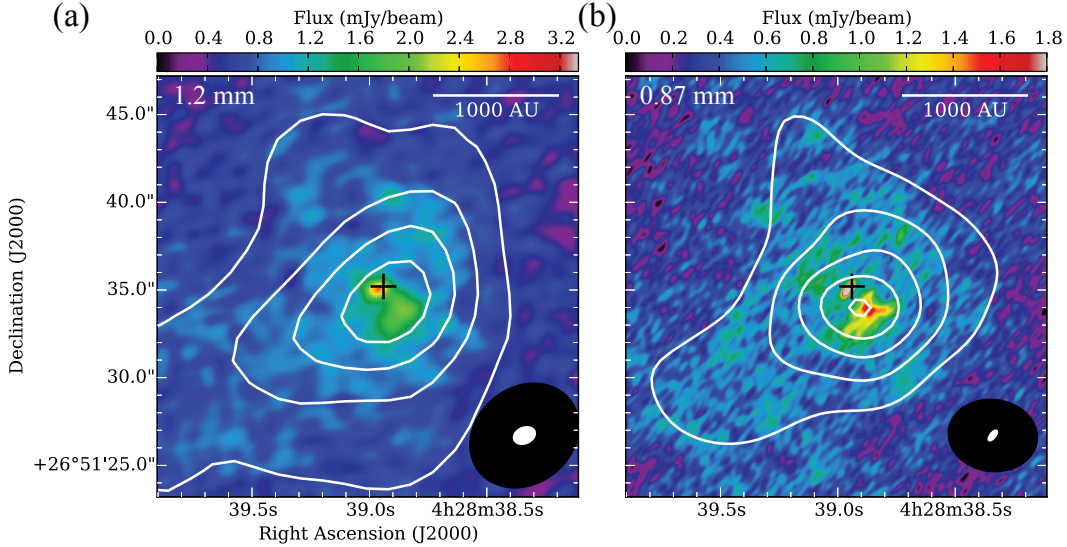}
 \end{center}
 \caption{Distributions of dust continuum emission toward MC27/L1521F. (a) Color scale image shows 1.2 mm dust continuum image combining the ALMA data (the 12 m array + the 7 m array) with IRAM 30 m data same as Figure \ref{Band6_Combine} (b). White contours illustrate 1.2 mm dust continuum emission taken by the 7 m array alone same as in Figure \ref{Band6_7m12m} (a). The angular resolutions of the combined image and the 7 m array alone are given by white and black ellipses, respectively, in the lower right corner. 
(b) The color scale image shows 0.87 mm dust continuum image combining the ALMA data (the 12 m array + the 7 m array) with JCMT/SCUBA data \citep{Shinnaga04,Kirk05}. White contours illustrate 0.87 mm dust continuum emission taken by the 7 m array alone. The contours start at three times the noise level and increase at this interval; the noise levels of the 7 m array data alone is 2.4 mJy beam$^{-1}$. The angular resolutions of the combined image and the 7 m array alone are given by white and black ellipses, respectively, in the lower right corner. 
Black crosses in each panel represent the position of the $Spitzer$ source.}
 \label{Band6_Band7_Combine}
\end{figure*}

\begin{figure*}
 \begin{center}
  \includegraphics[width=160mm]{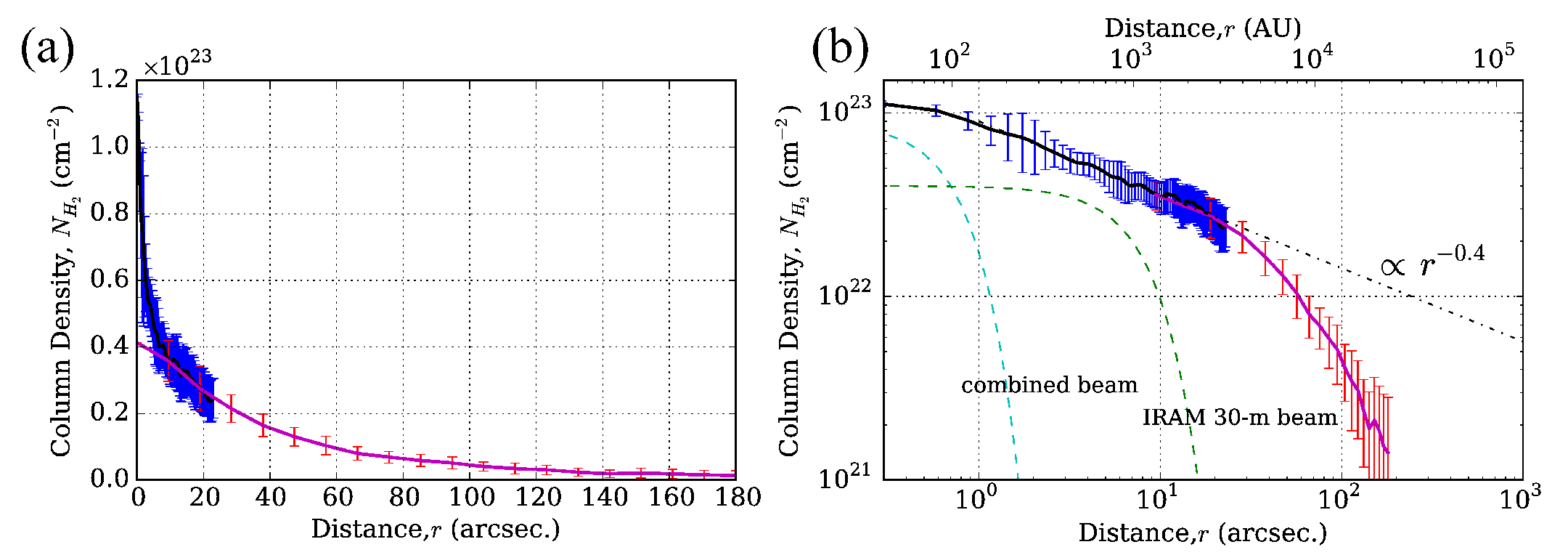}
 \end{center}
 \caption{Mean radial profile of H$_{2}$ column density centered at the MMS-2 peak in MC27/L1521F made from the 1.2 mm dust continuum image obtained by the IRAM 30 m telescope alone and the combined ALMA data (the 12 m array + the 7 m array) with the IRAM 30 m data. Panels (a) and (b) show the linear--linear plot and the log--log plot of the profiles, respectively. The averaged profiles of the combined data and the IRAM data are shown by black and magenta solid lines, respectively. Blue and red bars show the ($\pm$1$\sigma$) dispersion of the distribution of radial profiles in each data. The dot-dahed line in panel (b) denotes pure power-law distributions of $r^{-0.4}$.} Dotted curves in green and cyan in panel (b) indicate the beam shape of the IRAM 30 m telescope image and the combined image, a Gaussian function with FWHM of 14\arcsec and 1\farcs1, respectively.
 \label{power_low}
\end{figure*}

\begin{figure*}
 \begin{center}
  \includegraphics[width=160mm]{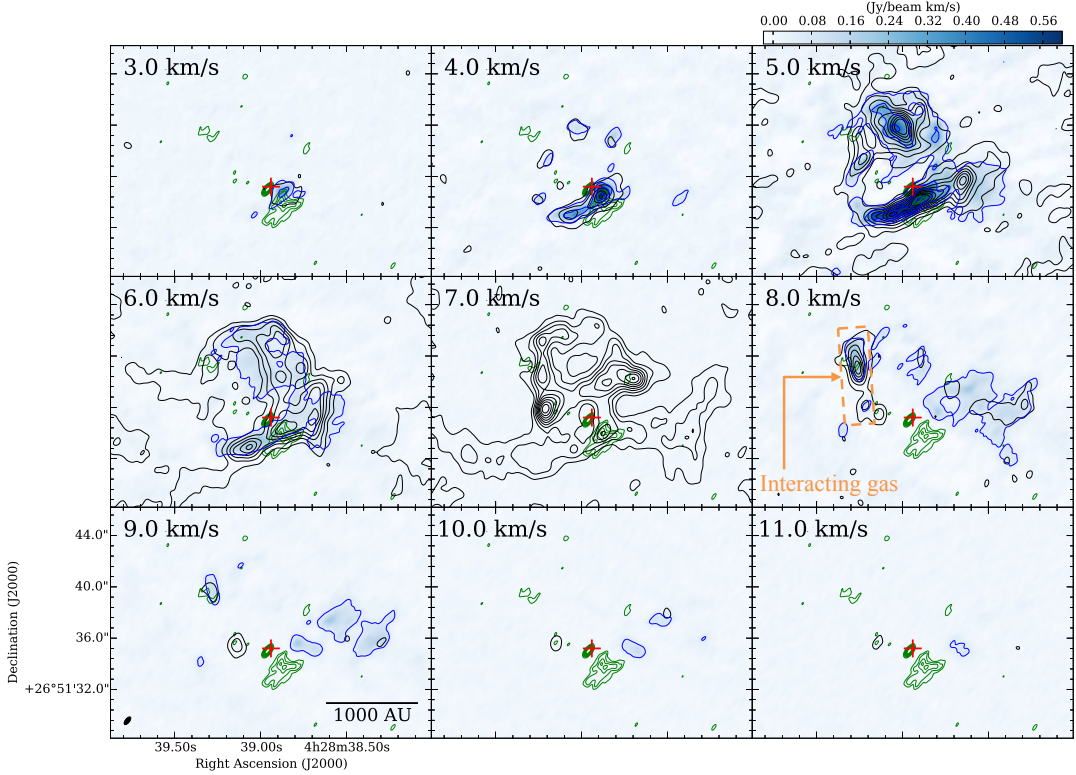}
 \end{center}
 \caption{Velocity-channel maps of the $^{12}$CO ($J$ = 3--2) and HCO$^{+}$ ($J$ = 3--2) emission toward MC27/L1521F. Bluecolor scale and blue contours show velocity-range-integrated intensity maps of $^{12}$CO ($J$ = 3--2) data. Black contours show those of HCO$^{+}$ ($J$ = 3--2) data \citep{Tokuda14}. The lowest contour and subsequent step of the blue contours are 0.05 Jy beam$^{-1}$ km s$^{-1}$ and 0.2  Jy beam$^{-1}$ km s$^{-1}$, respectively. The lowest contour and subsequent contour step of the black contours are 0.02 Jy beam$^{-1}$ km s$^{-1}$ and 0.04 Jy beam$^{-1}$ km s$^{-1}$, respectively. The velocity span for each map is 1.0 km s$^{-1}$. The lowest velocities are given in upper left corner of each panel. Green contours show the image of 0.87 mm dust continuum emission, as in Figure \ref{MMS2}. The angular resolution of the $^{12}$CO  ($J$ = 3--2) is given by the white ellipse in the lower left corner of the bottom left panel, 0\farcs73 $\times$ 0\farcs33. Red plus signs of each panel represent the position of the $Spitzer$ source. The dashed orange rectangle in panel of the 8.0 km\ s$^{-1}$ shows a interacting gas with a compact outflow, discussed by \cite{Tokuda14}.}
 \label{$^{12}$COchannel}
\end{figure*}

\begin{figure*}
 \begin{center}
  \includegraphics[width=160mm]{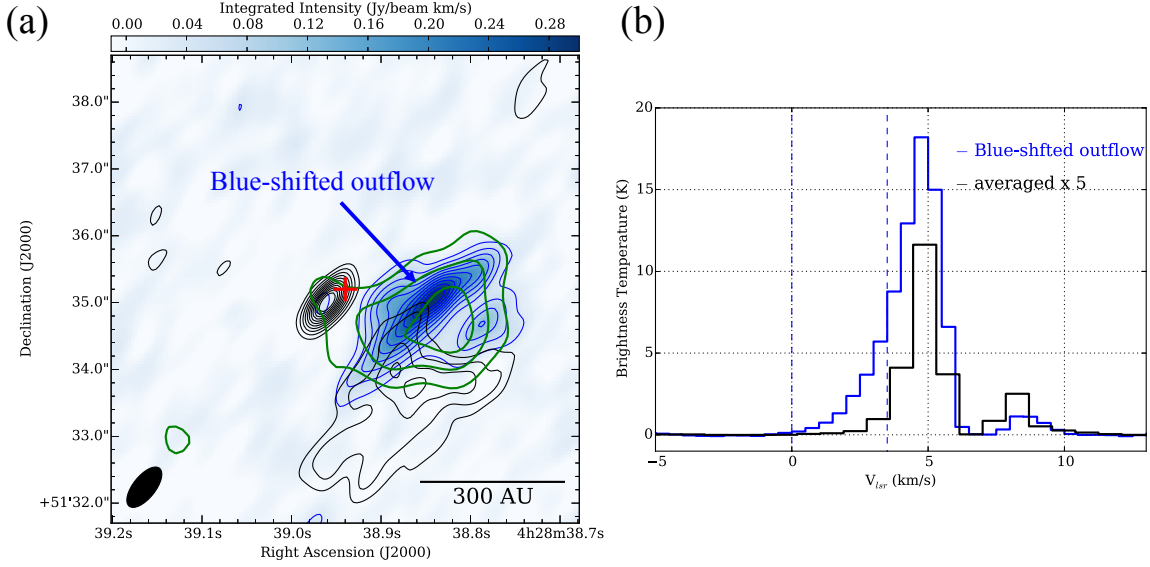}
 \end{center}
 \caption{Distribution of blueshifted outflow from the $Spitzer$ source. (a) The bluecolor scale image and blue contours show images of velocity-integrated intensity of $^{12}$CO ($J$ = 3--2) with a velocity range of 0.0--4.0 km$^{-1}$. The lowest contour and subsequent contour step are 0.03 Jy beam$^{-1}$ km s$^{-1}$.  Black contours show the image of 0.87 mm dust continuum emission, as in Figure \ref{MMS2}. The angular resolution of the $^{12}$CO ($J$ = 3--2) is given by the black ellipse in the lower left corner of the bottom left panel, 0\farcs73 $\times$ 0\farcs33. Red cross  represents the position of the $Spitzer$ source.  Green contours show images of velocity-integrated intensity of HCO$^{+}$ ($J$ = 3--2) with a range of 0.0--4.0 km$^{-1}$.  (b) Blue and black profiles show averaged spectra over the regions inside the blue lowest contour and a circle of with a radius of 9\farcs0 at the center position of ($\alpha_{J2000.0}$, $\delta_{J2000.0}$) = (4$^{\rm h}$28$^{\rm m}$38\fs80, +26\arcdeg51\arcmin38\farcs3), respectively. The blue dashed line shows the velocity range for blue contours in panel (a).}
\label{$^{12}$COoutflow}
\end{figure*}

\clearpage

\begin{deluxetable}
{lcccc}  
\tabletypesize{\scriptsize}
\tablecaption{Beam sizes and sensitivities \label{tablePram}}
\tablewidth{0pt}
\tablehead{
  & \multicolumn{2}{c}{Band 6 (1.2 mm)}
  & \multicolumn{2}{c}{Band 7 (0.87 mm)}\\
Parameters & 12 m array & 7 m array & 12 m array & 7 m array
}
\startdata
Synthesized beam size & $\sim$1\farcs3 $\times$ 0\farcs8 & $\sim$6\farcs6 $\times$ 5\farcs5  &  $\sim$0\farcs74 $\times$ 0\farcs32 & $\sim$5\farcs1 $\times$ 4\farcs2\\
Beam position angle$^{\rm a}$ & -66$\fdg$8 & -51$\fdg$6  & -38$\fdg$6 & 83$\fdg$1\\
Sensitivity of continuum observation (r.m.s.) & $\sim$0.1 mJy beam$^{-1}$ & $\sim$1.2 mJy beam$^{-1}$ & $\sim$0.1 mJy beam$^{-1}$ & $\sim$2.4 mJy beam$^{-1}$\\
Sensitivity of line observation (r.m.s.)$^{\rm b}$ & $\sim$7 mJy beam$^{-1}$ & $\sim$50 mJy beam$^{-1}$ & $\sim$5 mJy beam$^{-1}$ & $\sim$30 mJy beam$^{-1}$\\
\enddata
\tablenotetext{a}{The angle measured counterclockwise relative to the north celestial pole.}
\tablenotetext{b}{The velocity resolutions of the Band 6 and the Band 7 are $\sim$1.0 km\,s$^{-1}$ and $\sim$0.85 km\,s$^{-1}$, respectively.}
\end{deluxetable}

\begin{deluxetable}
{lccccccc}  
\tabletypesize{\scriptsize}
\tablecaption{Derived parameters of 0.87 mm dust sources \label{table1}}
\tablewidth{0pt}
\tablehead{
Source & $\alpha$  (J2000) & $\delta$  (J2000) &  $F_{\nu}$ (mJy)$^{\rm a}$ & Size (AU)$^{\rm b}$  & $F_\mathrm{max}$ (mJy beam$^{-1}$) & $N_\mathrm{max}$ (cm$^{-2}$)& $M_\mathrm{total}$ ($M_{\sun}$)
}
\startdata
MMS-1 & 4$^{\rm h}$28$^{\rm m}$38\fs96 & +26\arcdeg51\arcmin35\farcs0 & 2.6   & $\cdots$ & 2.7 & $\cdots$ & $\cdots$ \\
MMS-2 & 4$^{\rm h}$28$^{\rm m}$38\fs89 & +26\arcdeg51\arcmin33\farcs9 & 11.8  & 360 & 1.1 &1.3 $\times$ 10$^{23}$ & 3.9 $\times$ 10$^{-3}$\\
MMS-3 & 4$^{\rm h}$28$^{\rm m}$38\fs72 & +26\arcdeg51\arcmin38\farcs0 & 0.3    & 35   & 0.4 &5.1 $\times$ 10$^{22}$ & 1.8 $\times$ 10$^{-4}$\\
\enddata
\tablenotetext{a}{Flux of the dust emission integrated above the 3$\sigma$ level (1$\sigma$ rms $\sim$0.1 mJy beam$^{-1}$)}
\tablenotetext{b}{Diameter of a circle having the same area above the 3$\sigma$ after being deconvovled with the synthesized beam.}
\end{deluxetable}

\end{document}